 \documentclass[final,5p,times,twocolumn]{elsarticle}

 \usepackage{graphics,epsfig,subfigure}

\usepackage{amsmath}
\usepackage{amssymb}
\usepackage{cases}

\usepackage{amstext}
\usepackage{color}
\usepackage{array}
\usepackage{multirow}
\usepackage{slashed}
\newcommand\beq{\begin{equation}}
\newcommand\eeq{\end{equation}}
\newcommand\beqn{\begin{eqnarray}}
\newcommand\eeqn{\end{eqnarray}}
\newcommand\balign{\begin{align}}
\newcommand\ealign{\end{align}}
\newcommand\nn{\nonumber}
\newcommand\fc{\frac}
\newcommand\lt{\left}
\newcommand\rt{\right}
\newcommand\pt{\partial}
\allowdisplaybreaks
\biboptions{comma,square,compress,numbers,sort}

\journal{Physics Letters B}

\begin{document}

\begin{frontmatter}



\title{Braneworld sum rules and positive tension branes in a massive gravity}


\author[label1]{Ke Yang}
  \ead{keyang@swu.edu.cn}
\author[label2,label3]{Bao-Min Gu}
   \ead{gubm@ncu.edu.cn}
\author[label4,label5]{Yi Zhong\corref{cor1}}
  \ead{zhongy@hnu.edu.cn}
  \cortext[cor1]{The corresponding author.}

\address[label1]{School of Physical Science and Technology, Southwest University, Chongqing 400715, China}
\address[label2]{Department of Physics, Nanchang University, Nanchang 330031, China}
\address[label3]{Center for Relativistic Astrophysics and High Energy Physics, Nanchang University, Nanchang 330031, China}
\address[label4]{School of Physics and Electronics Science,
             Hunan University, Changsha 410082, China}
\address[label5]{Lanzhou Center for Theoretical Physics, Key Laboratory of Theoretical Physics of Gansu Province, and Key Laboratory of Quantum Theory and Applications of MoE, Lanzhou University, Lanzhou, Gansu 730000, China}

\begin{abstract}

By taking advantage of the braneworld sum rules, we explore the feasibility of constructing a flat 3-brane scenario consisting solely of positive tension branes in a 5D extension of the Lorentz-violating massive gravity. It is found that the theory supports three distinct brane configurations, one of which is exactly what we expected, consisting solely of two positive tension branes. The cosmological problem of Randall-Sundrum-1 model and the gauge hierarchy problem can be solved in this model simultaneously. Furthermore, the analysis of linear perturbations reveals that the tensor, vector and scalar modes are all massive and share the same mass spectrum, except that the ground state of vector mode is absent. Moreover, the tensor and vector modes are robust, but the scalar mode is ghost-like. Interestingly, even though the Kaluza-Klein gravitons have an extremely small mass splitting scale, an estimation of the effective gravitational potential and production of these gravitons on the brane indicates that the phenomenology of the present model is equivalent to that of the 6D ADD model.

\end{abstract}

\begin{keyword}
Extra dimensions \sep Randall-Sundrum model \sep Braneworld sum rules
\end{keyword}

\end{frontmatter}



\section{Introduction}

The possibility that our spacetime may have more dimensions other than four has always been concerned by theoretical physicists since the proposal of Klein-Kaluza (KK) theory in 1920s \cite{Kaluza1921,Klein1926}. Inspired by the string theory, the braneworld scenario is an available mechanism for hiding the undiscovered extra dimensions, where our visible universe is a 3-brane embedded in a higher-dimensional bulk and all  Standard Model particles are confined on the brane. In 1999, Randall and Sundrum proposed a well-known braneworld model, which provides a natural mechanism to solve the long-standing gauge hierarchy problem of particle physics \cite{Randall1999}. In the Randall-Sundrum-1 (RS1) model, there is a 3-brane located at each boundary of the orbifold extra dimension, where the one with negative tension is the infrared (IR) brane (or visible brane) our universe confined on, and the other with positive tension is the ultraviolet (UV) brane (or hidden brane).
 
It is well known that the overall sign of the source terms in the induced Friedmann-like equation on the brane depends on the sign of the brane tension \cite{Csaki1999,Cline1999,Shiromizu2000}. Since we live on a negative tension brane in RS1 model, it would lead to a ``wrong-signed" Friedmann-like equation and hence our observed expanding universe cannot be recovered on the brane. Moreover, it was found that the Standard Model fields can be localized on some positive tension branes, such as D-branes and NS-branes \cite{Lykken2000}. Hence, it is more reasonable to place our universe on a positive tension brane. One example of such a scenario is the Randall-Sundrum-2 (RS2) model, which features a single positive tension brane that we live on, but the gauge hierarchy problem is left \cite{Randall1999a}. In order to solve the gauge hierarchy problem and cosmological problem simultaneously, the authors considered some generations of RS1 model in modified gravitational theories \cite{Yang2012a,Yang2014,Guo2018}, where the massless 4D graviton is localized on the negative tension brane, so our world should move onto the positive tension brane in order to solve the gauge hierarchy problem. 

However, the negative tension brane is a potentially unstable object \cite{Cheng2010}, so it would be a better way to construct the braneworld scenario with only positive tension branes. In Ref.~\cite{Lykken2000}, the authors added a probe brane with a small positive tension into the RS2 model to build a hierarchy resolved configuration. For the exact brane solution, it is easy to show that the constraints require the presence of negative tension brane in a 5D compactification scheme in general relativity by resorting to the technique of braneworld sum rules \cite{Gibbons2001}.  However, it can be evaded for a higher-dimensional model such as the 6D anti-de Sitter soliton \cite{Leblond2001}. It is also found that the constraints can be relaxed in some modified gravitational theories even in 5D spacetime,  such as in scalar-tensor gravity \cite{Abdalla2008,Abdalla2010} and $f(R)$ gravity \cite{HoffdaSilva2011,Silva2017}. It gives us a hint that an extended RS1-like model with only positive tension branes could be realized in some modified gravitational theories.

In this work, we are interested in building braneworld model with only positive tension branes in a massive gravity, which is a generalization of general relativity by endowing the graviton with a nonzero mass. See Refs.~\cite{Hinterbichler2012,Rham2014,Rham2017} and the references therein for an introduction on massive gravity theories. Specifically, under the help of braneworld sum rules, we focus on generating the RS1-like scenario in a 5D extension of the Lorentz-violating massive gravity \cite{Lin2014}. In order to obtain a flat 3-brane configuration, 4D Poincar\'e invariance has to be preserved in this theory. Therefore, we assume that the background spacetime is invariant under 4D Poincar\'e transformation in the 5D extension of the Lorentz-violating massive gravity. The 5D diffeomorphisms are spontaneously broken due to the condensation of four background scalars. Then, the condensation generates four Goldstone excitations associated with the broken symmetries. Consequently,  the 5D massless spin-2 graviton with five degrees of freedom gets weight and possesses nine degrees of freedom on the spectrum by ``eating" the four Goldstone excitations in the unitary gauge. Other works related to braneworld scenario in massive gravities can be found in Refs.~\cite{Dvali2000,Dvali2000a,Chacko2004,Gabadadze2004,Kakushadze2014,Gabadadze2017,Gabadadze2018,Kaloper2019}.

 The layout of the paper is as follows: In Sect.~\ref{Sum_rules}, the constraint from the braneworld sum rules is discussed in the 5D extension of the Lorentz-violating massive gravity. In Sect.~\ref{Model}, a hierarchy-resolving toy model is built. The corresponding mass spectra of KK particles are discussed in Sect.~\ref{Perturbation}, and some low-energy phenomenology of the model is investigated in Sect.~\ref{Phenomenology}. Finally, brief conclusions and discussions are presented. Throughout the paper, the small Latin letters $(a, b, \cdots=0,1,2,3)$ are used to label the group indices of the internal metric of scalar fields, while the capital Latin letters $(A, B, \cdots=0,1,2,3,5)$ and Greek letters $(\mu,\nu, \cdots=0,1,2,3)$ are used to label the 5D and 4D spacetime indices, respectively.

\section{Braneworld sum rules}\label{Sum_rules}

To obtain a flat 3-brane configuration, we start from the most general metric ansatz keeping the four-dimensional Poincar\'e invariance, given by
\beq
ds^2=g_{MN}dx^M dx^N=a^2(y)\eta_{\mu\nu}(x)dx^\mu dx^\nu+dy^2,
\label{Brane_Metric}
\eeq 
where $a(y)$ is the warp factor,  and $y \in [-y_\pi, y_\pi]$ denotes a compact $S^1/Z_2$ orbifold extra dimension. Correspondingly, the 5D Ricci tensor can be written as
\begin{align}
R_{\mu\nu}=-a^2\eta_{\mu\nu}\lt(H'+4H^2\rt),\quad
R_{55}=-4\lt(H'+H^2\rt),
\end{align}
where the prime denotes the derivative with respect to $y$, and $H\equiv a'/a$. By tracing the above equations respectively, one obtains
\begin{align}
R^\mu_\mu  =-4\lt(H'+4H^2\rt), \quad
R^5_5 = -4\lt(H'+H^2 \rt).\label{Ricci_uu55}
\end{align}
Further, with the relation 
\beq
\lt(a^{\alpha+1} H\rt)'=a^{\alpha+1}\lt[H'+(\alpha+1)H^2 \rt],
\eeq
where $\alpha$ is an arbitrary constant, one obtains a useful relation by combining Eqs.~\eqref{Ricci_uu55}, i.e.,
\begin{align}
\lt(a^{\alpha+1} H\rt)'=\fc{a^{\alpha+1}}{4}\lt[\lt(\alpha-3\rt){R^5_5}-\alpha R^\mu_\mu\rt]. \label{Sum_Rule}
\end{align}

Here, we would like to consider a theory with a 5D Einstein-Hilbert term plus four canonical scalar fields, whose action is given by
\begin{align}
S&=M_*^3\int{d^5x}\sqrt{-g}\lt[\fc{R}{2}-\fc{1}{2}m^2g^{MN}\pt_M\phi^a\pt_N\phi^a - V(\phi^a\phi^a)\rt]\nn\\
&-\int d^4 x \sqrt{-g_{\text{I}}} V_{\text{I}}-\int d^4 x \sqrt{-g_{\text{II}}} V_{\text{II}},
\label{Main_Action}
\end{align}
where $M_*$ is the 5D fundamental gravity scale, $m$ is a parameter proportional to the mass of 5D graviton, $V(\phi^a\phi^a)$ is the self-interaction potential, and $V_\text{I}$ and $V_\text{II}$ represent the brane tensions at $y=0$ and $y=y_\pi$ respectively. In oder to achieve a Minkowski flat 3-brane model, the internal metric of the scalar fields has to be chosen as the Minkowski metric $\eta_{ab}$.

The corresponding field equations are obtained by varying the action (\ref{Main_Action}) with respect to the metric $g^{MN}$, 
\begin{align}
&R_{MN}-\fc{R}{2}g_{MN} = m^2 \Big(\partial_M \phi ^a \partial_N  \phi ^a - \frac{1}{2}  g_{MN} \partial^K \phi ^a \partial_K \phi ^a\Big) - V  g_{MN}\nn\\
&-\fc{V_{\text{I}}}{M_*^3 }g_{\text{I}\mu \nu} \delta ^{\mu }_M \delta ^{\nu }_N \delta  \left(y\right)-\fc{V_{\text{II}}}{M_*^3}g_{\text{II}\mu \nu } \delta ^{\mu }_M \delta ^{\nu }_N \delta  \left(y-y_\pi\right).\label{EoM}
\end{align}
Then, $R^\mu_\mu$ and $R^5_5$ can be expressed explicitly as
\begin{align}
R^\mu_\mu &= m^2 \partial^\mu \phi ^a \partial_\mu \phi ^a+\fc{8V}{3}+\fc{4V_{\text{I}}}{3M_*^3} \delta\left(y\right)+\fc{4V_{\text{II}}}{3M_*^3} \delta\left(y-y_{\pi}\right),\label{Ricci_uu_exp}\\
R^5_5 &=  m^2 ({\phi^a}')^2+\fc{2V}{3}+\fc{4V_{\text{I}}}{3M_*^3} \delta\left(y\right)+\fc{4V_{\text{II}}}{3M_*^3} \delta\left(y-y_{\pi}\right).\label{Ricci_55_exp}\
\end{align}
After inserting above Eqs.~\eqref{Ricci_uu_exp} and \eqref{Ricci_55_exp} into Eq.~\eqref{Sum_Rule}, one has
\begin{align}
&\lt(a^{\alpha +1}H \rt)'=-\fc{ a^{\alpha +1}}{4} \Big[\alpha  m^2 \partial^\mu \phi ^a\partial_\mu \phi ^a - (\alpha -3) m^2 ({\phi^a}')^2 \nn\\
& +2 (\alpha +1) V+4\frac{  V_\text{I}}{M_*^3}\delta (y)+4\frac{ V_{\text{II}}}{M_*^3} \delta  \lt(y-y_{\pi }\rt)\Big].
\label{Sum_rule_exp}
\end{align}
Since the $S^1/Z_2$ orbifold is periodic and compact, the integral of left hand side of the above equation \eqref{Sum_rule_exp} vanishes \cite{Gibbons2001}. Especially, for $\alpha=-1$, one obtains a useful constraint, 
\begin{align}
{M_*^3}\oint \Big( m^2 \partial^\mu \phi ^a\partial_\mu \phi ^a -4m^2  ({\phi^a}')^2\Big)dy = {4\lt(V_\text{I}+V_{\text{II}}\rt)}.
\label{Sum_rule_exp_2}
\end{align}
If the scalar fields further depend only on the extra dimension, which is typically the case considered in braneworld models, the constraint \eqref{Sum_rule_exp_2} reduces to
\beq
-m^2{M_*^3}\oint ({\phi^a}')^2dy =V_\text{I}+V_{\text{II}}.
\eeq
So in this case, the theory cannot support a model with only positive tension branes. Instead, if the scalar fields depend only on the brane coordinates, $\phi^a=\phi^a(x)$, the constraint \eqref{Sum_rule_exp_2} becomes
\beq
m^2{M_*^3}\oint \partial^\mu \phi ^a\partial_\mu \phi ^ady =4(V_\text{I}+V_{\text{II}}).
\eeq
It implies that the theory may support a model with only positive tension branes in this case. 

Specifically, here we consider a 5D extension of the Lorentz-violating massive gravity \cite{Lin2014}, where the background solution spontaneously breaks the 5D Lorentz invariance. The breaking of 5D Lorentz invariance stems from the condensation of scalar fields via 
\beq
\langle\phi^a\rangle=\delta_\mu^a x^\mu, 
\label{Scalar_vacuum}
\eeq 
with $x^\mu$ the brane coordinates. Thus, the condensation spontaneously generates a preferred 4D frame. Moreover, the scalar potential $V(\phi^a\phi^a)$ takes its vacuum value, i.e., the 5D cosmological constant $\Lambda$. In this case, the constraint simplifies to 
\beq
V_\text{I}+V_{\text{II}}=m^2{M_*^3}\oint a^{-2}dy.
\label{Tension_constraint}
\eeq

\section{Model building and hierarchy resolution}\label{Model}

With the ansatzes of the flat braneworld metric \eqref{Brane_Metric} and scalar field condensation \eqref{Scalar_vacuum}, the field equations \eqref{EoM} are written explicitly as
\begin{align}
3\lt(H'+2H^2 \rt)&=-\fc{m^2}{a^2}-\Lambda-\fc{V_\text{I} \delta\left(y\right)+ V_{\text{II}} \delta\left(y-y_{\text{b}}\right)}{M_*^3},\label{EoMII_1}\\
6H^2&=-\fc{2m^2}{a^2}-\Lambda. \label{EoMII_2}
\end{align}
From the field equations, we can easily obtain the solution
\beq
a(y)=e^{-k|y|}+\epsilon^2e^{k|y|}, 
\label{Warp_factor}
\eeq
where $\epsilon\equiv\fc{m}{2\sqrt{3}k}$, $k^2\equiv-\Lambda/6$, and $|y|$ represents the absolute value of $y$ in order to be consistent with the $Z_2$ symmetry. There are two branches in the solution of the warp factor, one is exponential growth and the other is exponential decay. Its minimum appears at $y_\text{m}=\log(1/\epsilon)/k$ with the value $a(y_\text{m})=2\epsilon$. 

By matching the delta functions in \eqref{EoMII_1}, we have the fine-tuning conditions,
\beq
V_\text{I}=6 k M_*^3 \left[1-\frac{2 \epsilon^2}{1+\epsilon^2}\right], ~~ V_\text{II}=6 k M_*^3 \left[1-\frac{2}{1+\epsilon^2 e^{2 k y_\pi}}\right].
\label{fine-tuning_conditions}
\eeq
It is straightforward to verify that the sum of brane tensions satisfies the constraint \eqref{Tension_constraint} from braneworld sum rules. 

\begin{figure}[htb]
\begin{center}
\subfigure[Configuration 1]  {\label{Model_1}
\includegraphics[width=3.7cm,height=3cm]{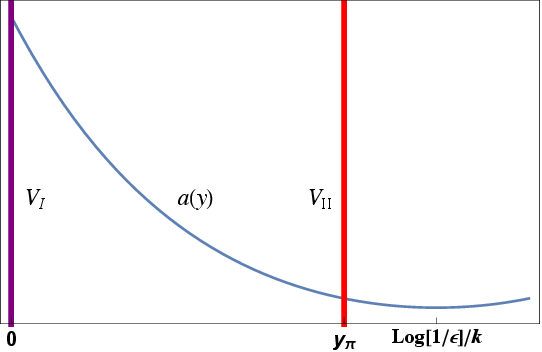}}\hfill
\subfigure[Configuration 2]  {\label{Model_2}
\includegraphics[width=3.7cm,height=3cm]{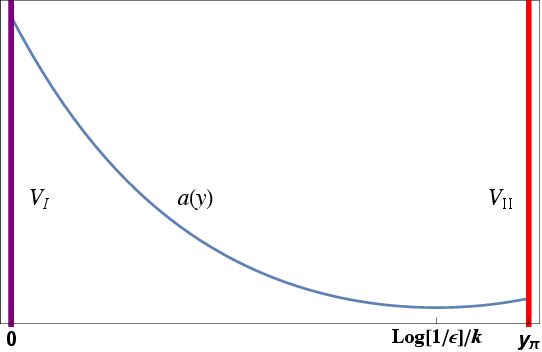}}
\end{center}
\caption{Diagrams of two distinct brane configurations.}
\label{Fig_Warp_factor}
\end{figure}

From the fine-tuning conditions (\ref{fine-tuning_conditions}), we observe that there are three different brane configurations depending on the size of extra dimension. Especially, the IR brane tension vanishes in the fine tuning conditions \eqref{fine-tuning_conditions} for $y_\pi=y_\text{m}$, so a single brane configuration is obtained in this case. Nevertheless, we are more interested in two brane configurations in this work, in which the gauge hierarchy problem may be solved. So we prefer to leave a brief discussion on it for the last section. Here we focus on the two brane configurations as illustrated in Fig.~\ref{Fig_Warp_factor}.

As shown in Fig.~\ref{Model_1}, the first configuration corresponds to the case of $e^{-k y_\pi}>\epsilon^2 e^{k y_\pi}$, i.e., the exponential decay branch of warp factor is dominant in the bulk. Then, the fine-tuning conditions (\ref{fine-tuning_conditions}) yield $V_I>0$ and $V_{II}<0$. Especially, in the limit $\epsilon e^{k y_\pi}\ll  1$, the exponential growth branch of the warp factor can be neglected compared to the exponential decay branch. Now the brane configuration reduces to that of RS1 model, with the fine-turning conditions,
\beq
V_I\approx-V_{II}\approx6 k M_*^3, ~~~ \Lambda=-6k^2.
\eeq
This brane configuration has been investigated in detail in the previous work by some of our authors \cite{Yang2022b}.

As shown in Fig.~\ref{Model_2}, the second configuration corresponds to the case of $e^{-k y_\pi}<\epsilon^2 e^{k y_\pi}<1$, where the exponential growth branch of warp factor is dominant near the IR brane. By observing from the fine-tuning conditions (\ref{fine-tuning_conditions}), it is interesting that both the brane tensions are positive in this case. This is the brane configuration that we have expected from the braneworld sum rules. Now since our universe is confined on the positive tension brane at $y_\pi$, a ``correct-signed" induced Friedmann-like equation can be obtained on the brane. 

As a crucial motivation of the braneworld, the gauge hierarchy problem can be solved in a natural way in this scenario. By confining the Higgs field on the IR brane at $y_\pi$, the fundamental Higgs vacuum expectation value (VEV) $v_0$ is redshifted by the warped factor, and therefore, the effective Higgs VEV measured by the observers on the brane is $v_\text{eff}=a(y_\pi) v_0$, which sets the electroweak scale of the Standard Model \cite{Randall1999}. If all the fundamental parameters $M_*$, $k$, $v_0$ are set to be the order of Planck scale $M_\text{Pl}\sim 10^{16}$TeV, there is no fundamental hierarchy between them. If the 5D graviton mass which is proportional to $m$ is light enough, i.e., $m/k<10^{-16}$, the minimum of warp factor $a(y_\text{m})=2\epsilon<10^{-16}$. Then, an effective TeV electroweak scale could be generated by just placing the IR brane at the place where $a(y_\pi)\sim10^{-16}$. 

We will see in the next section that the constraint on the 4D graviton mass implies $m<3.3\times 10^{-23}$eV, and as a result, the minimum of warp factor $a(y_\text{m})=2\epsilon\leq 10^{-51}$. Thus, the exponential growth branch of warp factor is completely dominant at IR brane, i.e., $a(y_\pi)\approx \epsilon^2 e^{k y_\pi}\sim10^{-16}$. Then, the size of the extra dimension is approximated given by $ y_\pi\approx \log\lt[{a(y_\pi)}/{\epsilon^2}\rt]/k$. For instance, $\epsilon\sim10^{-51}$ yields $y_\pi\approx 198/k$ and $\epsilon\sim10^{-60}$ yields $y_\pi\approx 239/k$. Therefore, the size of extra dimension is roughly two orders of magnitude larger than the Planck length. As a comparison, the size of extra dimension is $y_\pi\approx 37/k$ in RS1 model.

\section{Mass spectra of KK states}\label{Perturbation}

In order to investigate the mass spectra of KK excitations, we consider the linear perturbations against the background, 
\beq
ds^2=\lt(g_{MN}+h_{MN}\rt)dx^Mdx^N,
\label{Metric_NC}
\eeq
where $g_{MN}$ is the background metric \eqref{Brane_Metric} and $h_{MN}$ represents the linear perturbations. Due to the 4D Lorentz invariance of the background spacetime, it is convenient to decompose the perturbations $h_{MN}$ into the scalar, transverse vector and transverse-traceless tensor modes, as
\begin{align}
h_{55}&=-2\xi,\\
h_{\mu 5}&=-a \lt(S_{\mu}+\pt_\mu \beta \rt),\\
h_{\mu\nu}&=a^2\lt[ D_{\mu\nu}+2\eta_{\mu\nu}\psi+\fc{1}{2}\lt(\pt_{\mu}F_{\nu}+\pt_{\nu}F_{\mu} \rt)+2\pt_\mu\pt_\nu E \rt],
\end{align}
where the transverse-traceless tensor $D_{\mu\nu}$ satisfies the condition $\eta^{\mu\rho}\pt_\rho D_{\mu\nu}=0$ and the transverse vector modes $S_{\mu}$ and $F_{\mu}$ satisfy the condition $\eta^{\mu\rho}\pt_\rho S_{\mu}=\eta^{\mu\rho}\pt_\rho F_{\mu}=0$. 

The perturbed scalar fields are $\phi^a=x^a+\pi^a$, with $\pi^a=\delta^a_\mu\pi^\mu$ the Goldstone excitation of the condensation. The Goldstone excitation transforms like a vector field under the general coordinate transformation in order to maintain the scalar condensation \eqref{Scalar_vacuum} unchanged, i.e., $\pi^\mu\to\pi^\mu-\epsilon^\mu$. Correspondingly, it can be decomposed as $\pi^\mu=\eta^{\mu\nu}\lt(\pt_\nu \varphi+A_\nu \rt)$, where $\varphi$ is a scalar field and $A_\mu$ a transverse vector field satisfying $\eta^{\mu\rho}\pt_\rho A_{\mu}=0$.

Note that the quantity $Z_\mu-\pi_\mu$ is a gauge invariant quantity, where $Z_\mu\equiv a^2\lt(F_\mu/2+\pt_\mu E \rt)$ does not contain physical degrees of freedom in general relativity. In unitary gauge, the Goldstone excitations $\pi^\mu$ of scalar fields vanish. Then, it is clear that the four Goldstone excitations  $\pi^\mu$ are ``eaten" by $Z_\mu$, which survives in the linear perturbation theory and becomes physical degrees of freedom. Consequently, the 5D massless spin-2 graviton with five degrees of freedom gets weight and possesses nine degrees of freedom on the spectrum after ``eating" the four Goldstone excitations. 

By substituting the full perturbed metric \eqref{Metric_NC} into the action  \eqref{Main_Action} and expanding the action to the quadratic order of perturbations, the transverse-traceless tensor, transverse vector and scalar modes are decoupled with each other, so they can be treated separately. Working in the unitary gauge  $A_\alpha=\varphi=\psi=0$, the quadratic actions for tensor, vector and scalar modes are finally obtained respectively \cite{Yang2022b}, 
\begin{align}
S^{(2)}_\text{T} \!&=\!- \frac{M_*^3}{2}\!\int \! d^4 x dz a^3 \Big[\dot{\tilde{D }}_{\alpha \beta }\dot{\tilde{D }}^{\alpha \beta }\!+\!\partial ^{\alpha }\tilde D_{\alpha \beta }\partial_{\alpha }\tilde D^{\alpha \beta }\!+\!{2 m^2}\tilde D_{\alpha \beta }\tilde D^{\alpha \beta } \Big],\label{Tensor_action}\\
S^{(2)}_\text{V} \!&\!= -\frac{M_*^3}{2}\int d^4 x dz a^3  \Big[\dot{\tilde{F }}_{\alpha} \dot{\tilde{F }}^{\alpha}\!+\!\partial ^{\alpha }\tilde{F}_{\alpha }\partial_{\alpha } \tilde{F}^{\alpha }\!+\!{2 m^2}\tilde{F}_{\alpha } \tilde{F}^{\alpha }\Big],\label{Vector_action}\\
S^{(2)}_\text{S} \!&=\! \frac{M_*^3}{2}\int d^4 x dz a^3 \left[\dot{\tilde{E} }\dot{\tilde{E}} +\partial ^{\alpha }\tilde{E}\partial_{\alpha }\tilde{E}+{2 m^2}\tilde{E}\tilde{E}\right],
\label{Scalar_action}
\end{align}
where the indices are raised and lowered by the 4D Minkowski metric $\eta_{\mu\nu}$, $z$ is the conformal coordinate obtained through a coordinate transformation $dy=adz$, and in order to canonically  normalize these actions, the modes have been rescaled as $\tilde{D}_{\alpha \beta }=\fc{D_{\alpha \beta } }{2}$, $\tilde{F}_{\alpha }=\sqrt{\fc{k^2 m^2}{k^2+2 m^2}}\frac{ F_{\alpha }} {2}$, and $\tilde E= \sqrt{\frac{-3 k^4 m^2}{3 k^2+4 m^2}}E$, with $k^\alpha$ the four-momentum of various modes.

The forms of the three actions are similar, except that there is an overall wrong-sign in the action \eqref{Scalar_action} of scalar mode. Thus, the scalar perturbation $\tilde E$ is a ghost field. Further, by varying the actions \eqref{Tensor_action}, \eqref{Vector_action}, and \eqref{Scalar_action} with respect to various modes respectively, we have the equation of motion
\beq
\partial^{\alpha }\partial_{\alpha }\Upsilon+\ddot{\Upsilon}+3H\dot{\Upsilon }=2m^2{\Upsilon },
\label{EOM_Perturbations}
\eeq
where $\Upsilon$ represents $\tilde{D}_{\alpha \beta }$, $\tilde{F}_{\alpha }$, and $\tilde E$. With the KK decomposition $\Upsilon = \upsilon(x) a^{-\fc{3}{2}}(z)\Psi(z)$, the equation of motion \eqref{EOM_Perturbations} reduces to a 4D Klein-Gordon equations $\Box^{(4)}\upsilon(x)=M^2 \upsilon(x)$ and a Schr\"odinger-like equation,
\beq
-\ddot{\Psi} +\left(\frac{3}{2}\dot H+\frac{9}{4}H^2\right) {\Psi} =\mathcal{M}^2{\Psi} ,
\label{Sch_Eq}
\eeq
where $\mathcal{M}^2\equiv M^2-2m^2$, and $M$ is the effective mass of various KK states observed on the brane. 

The Hamiltonian can be factorized as $H_T=A_T^\dag A_T=\lt(\pt_z+\fc{3}{2}H\rt)\lt(-\pt_z+\fc{3}{2}H\rt)$, which is self-adjoint. With the Neumann boundary condition $\pt_z \tilde{D}_{ \alpha \beta } |_{z=0,z_b} =0$, it is easy to show that all the eigenvalues $\mathcal{M}^2$ are non-negative \cite{Yang2017}. Thus, it leads to $M^2\geq 2m^2$, namely, all the KK particles of various modes are massive. 

Further, by setting $\mathcal{M}^2=0$ or $M=\sqrt{2}m$, the ground state of the Schr\"odinger-like equation is obtained as
\beq
\Psi_{0}(z)=N_0a(z)^{\fc{3}{2}},
\eeq 
where the normalization factor $N_0$ can be worked out from the normalization condition $\int^{z_b}_{-z_b}\Psi_{0}^2dz=1$, yielding
\beqn
N_0^{-2}=\frac{1}{k}\lt[1-e^{-2 k y_\pi}+4\epsilon^2 k y_\pi -\epsilon^4(1-e^{2 k y_\pi})\rt].
\label{Normalization_factor}
\eeqn
Since the exponential growth branch of warp factor is completely dominant at the IR brane, the normalization factor is approximated as $N_0\approx \sqrt{k}$. 

It is noted that for the ground state of vector mode with the mass $M=\sqrt{2}m$, the rescaling $\tilde{F}_{\alpha }=\sqrt{\fc{k^2 m^2}{k^2+2 m^2}}\frac{ F_{\alpha }} {2}$ is ill-defined. Through a careful analysis, it is shown that the ground state of the vector mode does not exist in the mass spectra \cite{Yang2022b}. This is curial for recovering the mass spectra of RS1 model when the 5D graviton mass is turned off, as there is no massless vector mode in RS1 model due to the lack of continuous isometries of the bulk in the presence of 3-branes \cite{Randall1999}. Therefore, there exist only ground states of tensor and scalar modes in the mass spectra. 

From the KK decomposition $\Upsilon = \upsilon(x) a^{-\fc{3}{2}}(z)\Psi(z)$, the canonical normalized field configuration is given by $\Upsilon_{0} = \upsilon_{0}(x)$. Therefore, the lightest tensor and scalar modes propagate only on the brane, corresponding to the massive 4D graviton and massive radion respectively. However, the mass of 4D graviton is severely constrained by the gravitational experiments \cite{Rham2017}. For example, the detection of gravitational waves constrains the bound of the graviton mass to be $m_\text{g}\leq 4.7\times 10^{-23}$eV \cite{Abbott2019}. It leads to $m\leq3.3\times 10^{-23}$eV in our model. Under the constraint, the radion is also quasi-massless. However, the radion can gain weight through Goldberger-Wise mechanism and decoupled from the low-energy mass spectra \cite{Goldberger1999a}.

By including only the contribution of the quasi-massless graviton in the action \eqref{Main_Action}, the 4D effective gravitational mass scale $M_{\text{eff}}$ reads,
\beq
M_{\text{eff}}^2=M_*^3\int^{y_\pi}_{-y_\pi} a^2 dy={N_0^{-2}}{M_*^3}. \label{Mass_scale_relation}
\eeq
Therefore, the 4D effective gravitational mass scale is given by $M_{\text{eff}}^2\simeq M_*^3/k$, which is the order of Planck scale as expected.

For the excited KK states, their wave functions $\Psi_{M}$ can be worked out by solving the Schr\"odinger-like equation \eqref {Sch_Eq}. However, due to the complicated form of the effective potential, the Schr\"odinger-like equation cannot be solved directly in $z$ coordinate. By noting that the warp factor $a^\text{L}(y)=e^{-ky}$ is dominated in the region $0<ky<\log(1/\epsilon)$, while  $a^\text{R}(y)=\epsilon^2 e^{ky}$ is dominated in $\log(1/\epsilon)<ky<ky_\pi$, the  Schr\"odinger-like equation can be solved approximately in these two regions respectively, yielding
\begin{align}
\Psi^\text{L}_{M}(y) &= e^{\frac{k y}{2}} \lt[N_n J_2\left(\frac{M}{k}e^{k y}\right)+C_1 Y_2\left(\frac{M}{k}e^{k y}\right)\rt],\\
\Psi^\text{R}_{M} (y) &= e^{-\frac{k y}{2}} \lt[C_2J_2\left(\frac{M}{k \epsilon ^2}e^{-k y}\right)+C_3 Y_2\left(\frac{M}{k \epsilon ^2}e^{-k y}\right)\rt],\label{Psi_R}
\end{align}
where $N_n$ is the normalization factor, and $J_2$ and $Y_2$ are Bessel functions of order 2. After imposing the boundary condition $\pt_y \tilde{D}_{ \alpha \beta } |_{y=0,y_\pi} =0$, i.e., ${\Psi^\text{L}_{M}}'-\fc{3}{2}\fc{a^\text{L}{}'}{a^\text{L}}\Psi^\text{L}_{M}|_{y=0}=0$ and $\Psi^\text{R}_{M}{}'-\fc{3}{2}\fc{a^\text{R}{}'}{a^\text{R}}\Psi^\text{R}_{M}|_{y=y_\pi}=0$, one has 
\begin{align}
C_1=- \frac{J_1\left(\frac{M}{k}\right)}{Y_1\left(\frac{M}{k}\right)}N_n,\quad
C_3=-\frac{J_1\left(\frac{M}{k \epsilon ^2}e^{-ky_\pi}\right)}{Y_1\left(\frac{M}{k \epsilon ^2}e^{-k y_\pi}\right)}C_2 .
\end{align}

Since the terms of $J_2$ dominate near $ky\sim\log(1/\epsilon)$ in both $\Psi^\text{L}_{M}(y)$ and $\Psi^\text{R}_{M}(y)$, joining the two functions together at $ky_\text{m}=\log(1/\epsilon)$ requires that $C_2=N_n/\epsilon$ and $J_2\left(\frac{M}{k \epsilon }\right)=0$. This condition yields a discrete eigenvalue spectrum as $\mathcal{M}_{n}=x_nk \epsilon=\fc{x_n m}{2\sqrt{3}}$,
where $x_n$ satisfies $J_2\left(x_n\right)=0$, e.g. $x_1=5.136$,  $x_2=8.417$ and  $x_3=11.620$. Thus, the mass spectrum of excited KK states is given by
\beq
M_{n}=m\sqrt{2+\fc{x_n^2}{12}}.
\label{Mass_splitting}
\eeq
By utilizing the approximate formula of the zero point of  $J_2$, $x_n\approx \lt(n+\fc{3}{4}\rt)\pi$, the mass spectrum  \eqref{Mass_splitting} can be approximated as  $M_{n}\approx \fc{x_n m }{2\sqrt{3}} \approx \fc{n \pi m }{2\sqrt{3}}$, for $n\gg 1$. Thus, the mass splitting reads $\Delta M_{n}\approx \fc{\pi m}{2\sqrt{3}}$. 

On the other hand, the mass splitting scale of excited KK states can also be estimated from the fact that it is approximately quantized in units of inverse size of conformal extra dimension, i.e., $\Delta M_n \sim {1}/{z_\pi}$. From the coordinate transformation $dy=adz$, one has
\beq
z_\pi=\frac{1}{k \epsilon} \left[\text{arctan}\left(\epsilon e^{k y_\pi}\right)-\text{arctan} \left(\epsilon \right)\right],
\eeq
where the integral constant has been chosen so that $z(y=0)=0$. With $\epsilon\leq 10^{-51}$ and $ky_\pi\sim \mathcal{O}(200)$, the formula can be rewritten approximately as $z_\pi \approx \fc{\pi}{2\epsilon k}=\fc{\sqrt{3}\pi}{m}$. So the mass splitting scale approximately reads   $\Delta M_{n} \sim  \fc{m}{\sqrt{3}\pi}$, which  is the same magnitude as the previous result. Since $\Delta M_{n} \sim m<10^{-23}$eV, these massive KK excited states are extremely light in this model. 
 
\section{Phenomenology on the brane}\label{Phenomenology}

Since the mass splitting scale of KK gravitons is extremely small, an enormous amount of KK gravitons could be easily produced in accelerators, which may cause unacceptable large experimental signals. Therefore, it is necessary to check some low-energy phenomenology on the IR brane. 

First, we consider the correction to the Newtonian gravitational potential on the brane. Due to the extremely small mass splitting scale, the mass spectrum of KK gravitons can be approximated as continuous. The effective gravitational potential between two point-like sources of mass $m_1$ and $m_2$ separated by a distance $r$ on the brane takes the form \cite{Csaki2000,Lykken2000a} 
\beq
U(r)= -G_4\fc{ m_1 m_2}{r}\lt(1+\int dM \rho(M) e^{-Mr} \rt),
\eeq
where $\rho(M)$ is the relative density of states on the IR brane for excited KK states, defined by \cite{Lykken2000a}, $\rho(M)\equiv\fc{\lt|\Psi_{M}(y_\pi) \rt|^2}{\lt|\Psi_{0}(y_\pi) \rt|^2}$. 

Since $\sqrt{x}J_2(x)\approx \sqrt{\fc{2}{\pi}}\cos\lt(x -\fc{5}{4}\pi\rt)$ for a large $x$, the dominant terms $J_2$ in $\Psi^\text{L}_{M}(y)$ and $\Psi^\text{R}_{M}(y)$  have the plane wave behavior around $y_\text{m}=\log(1/\epsilon)/k$. So after normalizing these KK states as plane waves such that the physical quantities always involve an integration over $M$ for which the proper measure is $dM$ \cite{Lykken2000,Lykken2000a}, the normalization factor is given by $N_n=\sqrt{{M}/{k}}$. Thus, from Eq.~\eqref{Psi_R}, we have  
\beq
\Psi_{M}(y_\pi)=\Psi^\text{R}_{M} (y_\pi)\approx -\lt(\fc{M}{k \epsilon^2 e^{k y_\pi}}\rt)^\fc{1}{2}.
\eeq
With $\Psi_{(0)}(y_\pi)\approx\sqrt{k} \left(\epsilon ^2 e^{k y_\pi}\right)^{3/2}$, the relative density of states is obtained finally
\beq
\rho(M)\approx \frac{M }{k^2 \lt(\epsilon^2 e^{k y_\pi}\rt)^4}.
\eeq
As a result, the effective gravitational potential on the IR brane reads
\begin{align}
U(r)&= -G_4\fc{ m_1 m_2}{r}\lt(1+ \frac{1}{k^2 r^2 \lt(\epsilon^2 e^{k y_\pi}\rt)^4} \rt)\nn\\
&=-G_4\fc{ m_1 m_2}{r}\lt[1+ \frac{1}{\lt(10^{-4}\text{eV}\rt)^2 r^2 }\rt].
\label{Gravitational_Potential}
\end{align}
This deviation is the same as that of ADD model with two extra dimensions \cite{Arkani-Hamed1998}.

Another widely considered process is the real emission of the KK gravitons, which could be observed as missing energy in the accelerators. The total cross section for the production of these on-shell massive gravitons in a typical process $e^+e^-\to\gamma+$ KK gravitons can be roughly estimated to be \cite{Lykken2000a}
\beq
\sigma(e^+e^-\to\gamma+\slashed{E}_\text{KK}) \sim \fc{\alpha}{M^2_\text{Pl}}\int_0^{E_c} dM \rho(M),
\eeq
where $E_c$ is the center of mass energy for the process and $\slashed{E}_\text{KK}$ is the missing energy carried away by KK gravitons. After some simple algebra, the final result for total cross section is of order
\beq
\sigma(e^+e^-\to\gamma+\slashed{E}_\text{KK})\sim \frac{\alpha E_c^2 }{ k^4 \lt(\epsilon^2 e^{k y_\pi}\rt)^4}
\sim \fc{\alpha E_c^2}{\text{TeV}^4}.
\eeq
This result is consistent with that of 6D ADD model as well \cite{Rubakov2001}. 

The reason why the phenomenology generated by only one extra dimension in present model are similar to those of 6D ADD model can be seen from the couplings of the excited KK gravitons to matter \cite{Davoudiasl2000,Yang2012a,Guo2018}, 
\begin{align}
{\zeta}_{n}\sim \frac{a^{-\fc{3}{2}}(y_\pi)\tilde\Psi_{M}(y_\pi)}{ M_*^{\fc{3}{2}}} \sim \frac{\sqrt{m M_n}}{\sqrt{k} M_\text{Pl}^{3/2} \left(\epsilon ^2 e^{k y_\pi}\right)^2}\sim \fc{\sqrt{n}m}{\text{TeV}^2}.
\end{align}
Here, $\tilde\Psi_{M}(y_\pi)\equiv\Psi_{M}(y_\pi)/\sqrt{z_\pi}$, in which,  the factor $1/\sqrt{z_\pi}$ is included in order to restore an appropriate dimension when returning from continuous integration to discrete summation \cite{Randall1999a}. It is clear that the coupling between excited KK gravitons to matter in present model ${\zeta}_{n}\sim \fc{m}{10^{-4}\text{eV}}\fc{\sqrt{n}}{M_\text{Pl}}$ is much smaller than that of ADD model in which ${\zeta}_{n}\sim 1/M_\text{Pl}$. Nevertheless, in present model, the number of species of KK gravitons with masses below $E_c$ is $n\sim (E_c z_\pi) \sim \fc{E_c}{\Delta M_n}\sim \fc{E_c}{m}$, which is much more than the number of 6D ADD model $n\sim(E_c R)^2\sim (\fc{E_c}{10^{-4}\text{eV}})^2$. Therefore,  from $\sigma(e^+e^-\to\gamma+\slashed{E}_\text{KK})\sim \sum_{n}\alpha{\zeta}_{n}^2\sim \alpha n {\zeta}_{n}^2\sim \fc{\alpha E_c^2}{\text{TeV}^4}$, the total cross sections of the two model are ultimately the same. 


\section{Conclusions and discussions}\label{Conclusions}

In RS1 model, the fact that our world resides on the negative tension brane results in a ``wrong-signed" Friedmann-like equation, making it unable to describe an expanding universe on the brane. To overcome this issue, one possible solution is to construct a braneworld scenario using only positive tension branes. By taking advantage of braneworld sum rules, it was found that this brane configuration may be supported in the 5D extension of the Lorentz-violating massive gravity. Therefore, we generalized RS1-like model in this gravity and found that the theory supports three distinct brane configurations, among which the configuration with only two positive tension branes is exactly what we expected. By confining our world on the positive tension IR brane, a ``correct-signed" 4D Friedmann-like equation can be recovered in this model and the gauge hierarchy problem can be solved as well.

By considering the full linear perturbations against the background metric, it is found that the tensor and vector modes are robust, but the scalar mode is ghost-like. By solving the quadratic actions of various perturbed modes, it was found that they share the same mass spectrum, which start from $\sqrt{2}m$ and have a mass splitting scale roughly $\Delta M_{n}\sim m$, except that the ground state of vector mode is absent in the mass spectrum. Since the graviton mass is severely constrained by experimental observations, the mass splitting of KK gravitons has to be extremely small. By further studying the interaction between KK gravitons and matter fields on the brane, it was found that the large amount of KK gravitons leads to the same phenomenology as that of the ADD model with two extra dimensions. 
 
If we remove the IR brane, i.e., $y_\pi\to\infty$, the ground state would no longer be normalizable. In this case, the effective 4D gravity cannot be recovered on the brane. So the RS2-like single brane model with non-compact extra dimension is not viable in present theory. However, it is interesting that a single brane model with compact extra dimension can be obtained by setting $y_\pi=y_\text{m}=\log(1/\epsilon)/k$ to vanish the IR brane tension in the fine tuning conditions \eqref{fine-tuning_conditions}. Now our world should move to the only remaining UV brane and hence the gauge hierarchy problem is left. The mass splitting scale approximates $\Delta M_{n}\sim {1}/{z_\pi}\sim \frac{2 m}{\sqrt{3} \pi }$, which is the same order as that of above model with positive tensions. However, due to the wave functions of KK particles are heavily suppressed by the potential barrier of order of  $M_\text{Pl}$ at $y=0$, the total cross section of the process $e^+e^-\to\gamma+$ KK gravitons is estimated as $\sigma\sim {E_c^2}/{M_\text{Pl}^4}$, and the gravitational potential reads $U(r)= -G_4\fc{ m_1 m_2}{r}\lt(1+ \frac{1}{k^2 r^2} \rt)$, which is in fact identical to those in the RS2 model.

In order to obtain a Minkowski flat brane model, the internal metric has been chosen as the Minkowski metric in the action \eqref{Main_Action}. However, due to the unbroken 4D Lorentz invariance and the non-Fierz-Pauli form of the mass term,  the ghost scalar mode appears at the linear perturbation theory. The ghost mode could be removed by breaking further the 4D Lorentz invariance on the brane, such as leaving a residual SO(3) symmetry. However, since the 4D Lorentz invariance is necessary to build the flat 3-brane configuration, there would be no flat brane solution anymore in a 4D Lorentz symmetry breaking theory. Therefore, one can only expect (anti)-de Sitter 3-brane solutions in such a theory with residual SO(3) symmetry. These models are left for our further investigation.

\section*{ACKNOWLEDGMENTS}

This work was supported by the National Natural Science Foundation of China under Grant Nos.~12005174, 12165013, and 12247101. K. Yang acknowledges the support of Natural Science Foundation of Chongqing, China (Grant No.~cstc2020jcyj-msxmX0370). B.-M. Gu acknowledges the support of Jiangxi Provincial Natural Science Foundation (Grant No.~20224BAB211026). Y. Zhong acknowledges the support of the Fundamental Research Funds for the Central Universities (Grant No.~531118010195) and the Natural Science Foundation of Hunan Province, China (Grant No.~2022JJ40033).


\end{document}